\documentclass[twocolumn]{aastex63}

\usepackage{epsfig}
\usepackage{url}
\usepackage{hyperref}

\usepackage[normalem]{ulem}

\usepackage{latexsym}
\usepackage{epsfig}
\usepackage{amsmath}
\usepackage{amssymb}
\usepackage{wasysym}
\usepackage{graphicx}
\usepackage{verbatim}
\usepackage{enumerate,mdwlist}
\usepackage{amsfonts}
\usepackage{pgfplots}
\usepackage{bbold}
\usepackage[normalem]{ulem}

\newcommand{\lb}{\left[}
\newcommand{\rb}{\right]}

\newcommand{\ba}{\begin{eqnarray}}
\newcommand{\ea}{\end{eqnarray}}
\newcommand{\be}{\begin{equation}}
\newcommand{\ee}{\end{equation}}

\newcommand{\dd}{\mathrm{d}}
\newcommand{\yr}{\mathrm{yr}}
\newcommand{\obs}{\mathrm{obs}}
\newcommand{\mmax}{m_\mathrm{max}}
\newcommand{\mmin}{m_\mathrm{min}}
\newcommand{\detc}{\mathrm{det}}
\newcommand{\Gpc}{\mathrm{Gpc}}

\newcommand{\Msun}{M_\odot}
\newcommand{\Mtot}{M_\mathrm{tot}}
\newcommand{\Mchirp}{\mathcal{M}_c}
\newcommand{\Mz}{\mathcal{M}_z}
\newcommand{\R}{\mathcal{R}}

\newcommand{\VTsen}{\langle VT\rangle_\mathrm{sen}}
\newcommand{\Ogw}{\Omega_\mathrm{gw}}

\definecolor{grey}{rgb}{0.4,0.4,0.4}
\definecolor{dullmagenta}{rgb}{0.4,0,0.4}
\definecolor{darkblue}{rgb}{0,0,0.4}
\definecolor{midblue}{rgb}{0,0,0.5}
\definecolor{midred}{rgb}{0.5,0,0}
\definecolor{orange}{rgb}{1,0.5,0}
\definecolor{lightbrown}{rgb}{0.75,0.5,0.25}
\definecolor{tan}{cmyk}{0.14,0.42,0.56,0}
\definecolor{djunglegreen}{cmyk}{0.99,0,0.52,0}
\definecolor{lightgreen}{rgb}{0,1,0}
\definecolor{olivegreen}{cmyk}{0.64,0,0.95,0.40}
\definecolor{midgreen}{rgb}{0.0,0.675,0.0}
\definecolor{darkgreen}{rgb}{0,0.5,0}

\usepackage[all]{xy} 
\usepackage{amsfonts}

\begin{document}

\title{Jumping the gap: searching for LIGO's biggest black holes}

\author{Jose Mar\'ia Ezquiaga}
\altaffiliation{NASA Einstein fellow}
\correspondingauthor{Jose Mar\'ia Ezquiaga}
\email{ezquiaga@uchicago.edu}
\affiliation{Kavli Institute for Cosmological Physics and Enrico Fermi Institute, The University of Chicago, Chicago, IL 60637, USA}

\author{Daniel E. Holz}
\affiliation{Kavli Institute for Cosmological Physics and Enrico Fermi Institute, The University of Chicago, Chicago, IL 60637, USA}
\affiliation{Department of Physics, Department of Astronomy \& Astrophysics, The University of Chicago, Chicago, IL 60637, USA}

\begin{abstract}
Gravitational wave (GW) detections of binary black holes (BBHs) have shown evidence for a dearth of component black holes with masses above $\sim50M_\odot$.
This is consistent with expectations of a mass gap due to the existence of pair-instability supernovae (PISN). We argue that ground-based GW detectors will be sensitive to BBHs with masses above this  gap, $\gtrsim120\,M_\odot$. With no detections, two years at upgraded sensitivity (A+) would constrain the local merger rate of these BBHs on the ``far side'' of the PISN gap to be lower than $0.01\,\mathrm{yr}^{-1}\mathrm{Gpc}^{-3}$. Alternatively, with a few tens of events we could constrain the location of the upper edge of the gap to the percent level. We consider the potential impact of ``interloper'' black holes within the PISN mass gap on this measurement. Far side BBHs would also be observed by future instruments such as Cosmic Explorer (CE), Einstein Telescope (ET) and {\it LISA}, and may dominate the fraction of multi-band events. We show that by comparing observations from ground and space it is possible to constrain the merger rate history.  Moreover, we find that the upper edge of the PISN mass gap leaves an imprint on the spectral shape of the stochastic background of unresolved binaries, which may be accessible with A+ sensitivity. Finally, we show that by exploiting the upper edge of the gap, these high-mass BBHs can be used as standard sirens to constrain the cosmic expansion at redshifts of $\sim0.4$, $0.8$, and~$1.5$ with {\em LISA}, LIGO-Virgo, and CE/ET, respectively. These far-side binaries would be the most massive BBHs LIGO-Virgo could detect.
\end{abstract}

\section{{Introduction}}
The first three observing runs of Advanced LIGO \citep{Aasi2015} and Virgo \citep{Acernese_2014} have shown evidence for a population of stellar-mass binary black holes (BBHs) with a dearth of masses above $\sim50\,\Msun$ \citep{Fishbach:2017zga,GWTC-1,LIGOScientific:2018jsj,Abbott:2020gyp}. This is consistent with pair-instability supernova (PISN) \citep{Barkat:1967zz,Fowler:1964zz,Heger:2001cd,Fryer_2001,Heger_2003,2016A&A...594A..97B}, a runaway process induced by electron-positron pair production occurring in massive stars. These PISN result in complete disruption of the stars, preventing the formation of remnant black holes and thus inducing a gap in the mass spectrum. However, for sufficiently massive stars the PISN process is insufficient to prevent direct collapse, and a population of intermediate mass black holes (IMBHs) with masses above $\sim120M_\odot$ is expected to arise. IMBHs have been a long-standing target of the LIGO Virgo Collaboration (LVC) \citep{Virgo:2012aa,Aasi:2014iwa,Salemi:2019ovz}, and firm upper bounds have been set on their merger rate with O1-O2 data \citep{Salemi:2019ovz}.

These ``far side" or ``post gap'' massive binaries appear in a range of formation channels. For example they may be produced in field binaries~\citep{Madau_2001,Belczynski:2014iua,refId0,Mangiagli_2019} or in globular clusters \citep{10.1093/mnras/stv2162,Rodriguez:2016kxx}.
Still, the PISN gap may not be strictly empty, since it could be partially filled with black holes from second generation mergers in dense star clusters \citep{2017ApJ...840L..24F,Gerosa:2017kvu,Gerosa:2019zmo,Rodriguez:2019huv} and galactic nuclei \citep{PhysRevLett.123.181101}, stellar collisions \citep{2019arXiv191101434D}, quadruple systems \citep{Fragione_2020}, or gas accretion \citep{refId00}.
However, existing data show that only $2^{+3.4}_{-1.7}\%$ of BBH systems 
have primary masses above $45\Msun$~\citep{Abbott:2020gyp},
robustly demonstrating a precipitous drop in the BBH population above $45\Msun$.
The fact that the population shows this drop is in some sense a strong validation of theoretical predictions of a PISN feature at precisely this location in the mass distribution. 

The LVC has recently announced GW190521 \citep{Abbott:2020tfl}, the most massive binary detected thus far. Under uninformative priors, GW190521 shows evidence for both the primary and secondary masses being located within the PISN gap \citep{Abbott:2020mjq}.
By assuming a population informed prior in which the secondary mass is considered a member of the O1+O2 population distributions, \citet{Fishbach:2020qag} have shown that the primary mass could instead be the first black hole on the far side of the PISN gap. Of course, this is not a definitive determination, but it further motivates our consideration of a far-side, post-PISN population.

BBHs above the PISN mass gap would be extraordinarily loud sources of GWs. Remarkably, these binaries lie at the intersection of ground- and space-based detector sensitivities, complementing stellar-mass multiband binaries \citep{Sesana:2016ljz}.
In this \emph{Letter} we study the sensitivity of present and future detectors to the mergers of far-side binaries.
We explore a range of gap locations and population properties, as well as different instruments.
We address the precision with which the end of the gap could be constrained, both from the detection of individual sources and from the stochastic background of unresolved sources. We also propose the use of the edge of the mass gap to directly constrain the cosmic expansion rate using standard sirens at high redshift.
Detecting this new population of black holes would have important implications for astronomy, fundamental physics, and cosmology.

The potential existence of ``interloper'' binaries within the gap does not invalidate the existence of a PISN mass gap. These may be members of a sub-population due to a completely different formation channel, such as hierarchical mergers.
Nevertheless, as we discuss further below, current observations allow for a significant population of BBHs {\em above}\/ the gap.
Despite evidence for BBHs at higher masses, existing data robustly demonstrates the presence of a sharp drop in BBHs with component masses $\gtrsim45\,\Msun$ \citep{Fishbach:2017zga,2020arXiv201014533T}.
This dearth of BBHs above $\sim 45\,\Msun$ is consistent with the presence of the lower edge of the PISN gap.
Similarly, current data is completely consistent with the existence of the existence of the upper edge of the PISN gap.
Because the sensitivity of the LIGO/Virgo detectors falters at higher mass \citep{Chen:2017wpg}, current data does not significantly constrain the population near the upper edge of the gap. In what follows, we will assume the existence of a ``far side'' population, consistent with theory and with existing upper limits. Furthermore, we will assume that any putative sub-population of BBHs within the gap is sub-dominant to the far side population above the upper edge of the gap (see Appendix~\ref{app:in_gap_pop}).
It is of course conceivable that no far-side BHs will be detected, or that the detected high-mass black holes are inconsistent with PISN theory and do not show the presence of the upper edge of the PISN gap. As shown below, data in the coming years will sharply constrain the existence or absence of a population of far side BHs.

\section{{LIGO-Virgo sensitivity above the PISN mass gap.}}
In order to determine the sensitivity of a GW interferometer to BBHs in a given mass range, it is useful to estimate the sensitive volume weighted by the observation time, $T_\obs$ \citep{Chen:2017wpg,Fishbach:2017zga}:
\be \label{eq:VTsen}
\VTsen(m_1,m_2)=T_\obs\int \dd z\frac{1}{1+z}\frac{\dd V_c}{\dd z}p_\detc(z,m_1,m_2)\,,
\ee
where $p_\detc$ encodes the probability of detecting a binary with source masses $m_1$ and $m_2$ at redshift $z$, and encapsulates details of the detector including
the power spectral density and antenna power pattern~\citep{2015ApJ...806..263D}.
Further details on the methodology and detector sensitivities are placed in the Appendix \ref{app:methods}.

\begin{figure}[t!]
\centering
\includegraphics[width = 0.95\columnwidth]{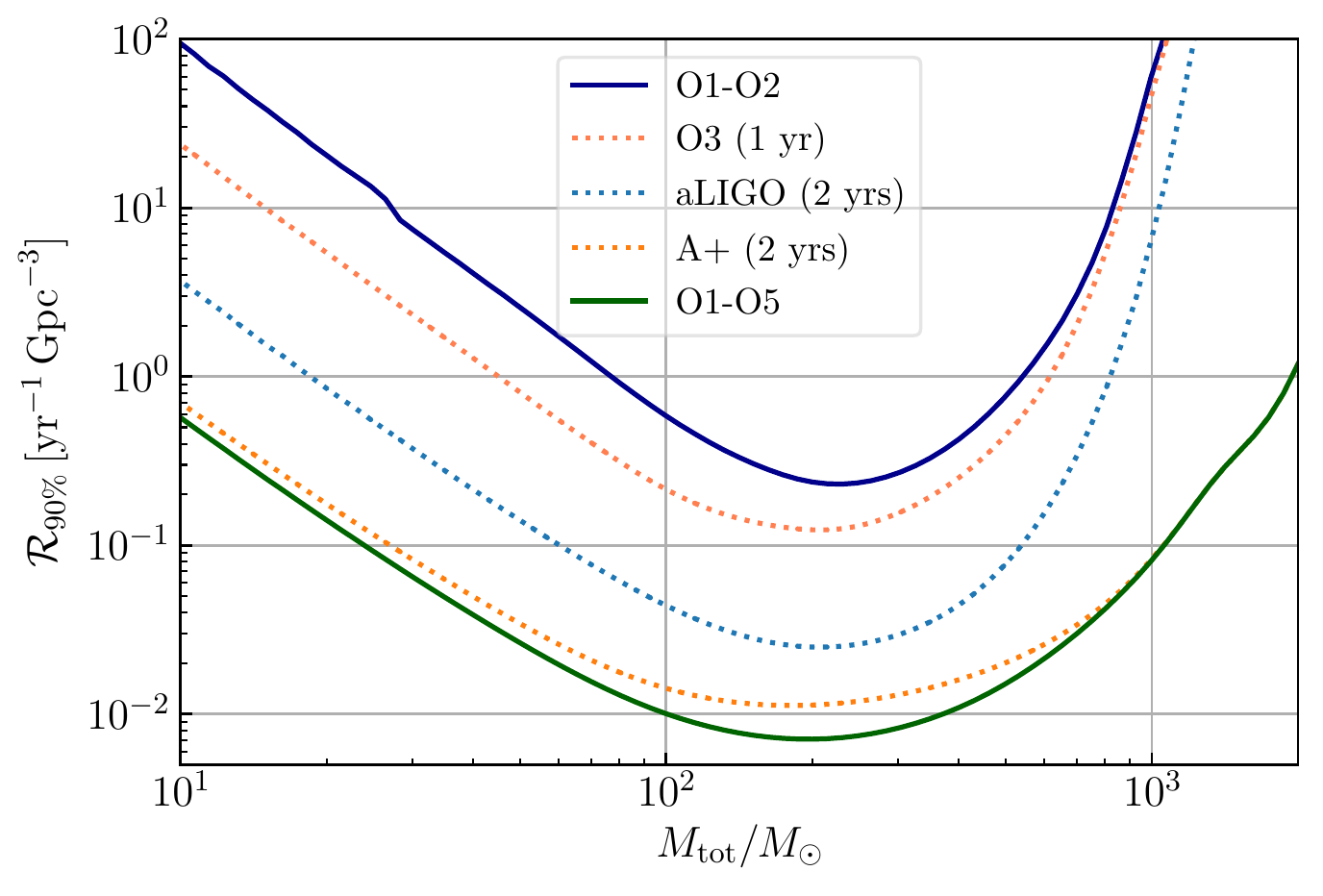}
\caption{Projected upper bounds on the merger rate in the \emph{absence}\/ of detections (90\% confidence upper limits) of past and future LIGO-Virgo observing runs as a function of the total mass.}
 \label{fig:upper_bound_rate}
\end{figure}

In the absence of detections, one can use $\VTsen$ to place an upper bound on the comoving merger rate, $\R$. Following the LVC-IMBH search analyses \citep{Salemi:2019ovz}, we estimate the 90$\%$ confidence upper limit for a given total mass as $\R_{90\%}=-\ln(0.1)/\VTsen$.  The corresponding projected upper bounds from the first two observing runs and a number of future campaigns are shown in Fig.~\ref{fig:upper_bound_rate}.
 We find that after two years at design sensitivity (aLIGO) the merger rate could be constrained to be below $0.03\yr^{-1}\Gpc^{-3}$ for a total mass within $100$--$200\Msun$. After two additional years at upgraded sensitivity (A+), the upper bound would be lowered to $0.01\yr^{-1}\Gpc^{-3}$.

Future LVC observing runs will be increasingly sensitive to far-side binaries. The number of detections will depend on the probability density function of the source masses, $p(m_1,m_2)$ and the redshift evolution of the merger rate, $\R(z)$:
\be
\frac{\dd^4N_\text{det}}{\dd m_1\dd m_2\dd z\dd t_d}=\frac{\R(z)}{1+z}\frac{\dd V_c}{\dd z}p(m_1,m_2)p_\text{det}\,.
\ee
We begin by considering the sensitivity as a function of the total mass of the binary, assuming equal mass components (we will explore BBH populations later). For the redshift evolution of $\R(z)$ we consider two representative scenarios: \emph{(i)} a constant merger rate, and \emph{(ii)} a merger rate following the star formation rate (SFR).

\begin{figure*}[t!]
\centering
\includegraphics[width = 0.98\textwidth]{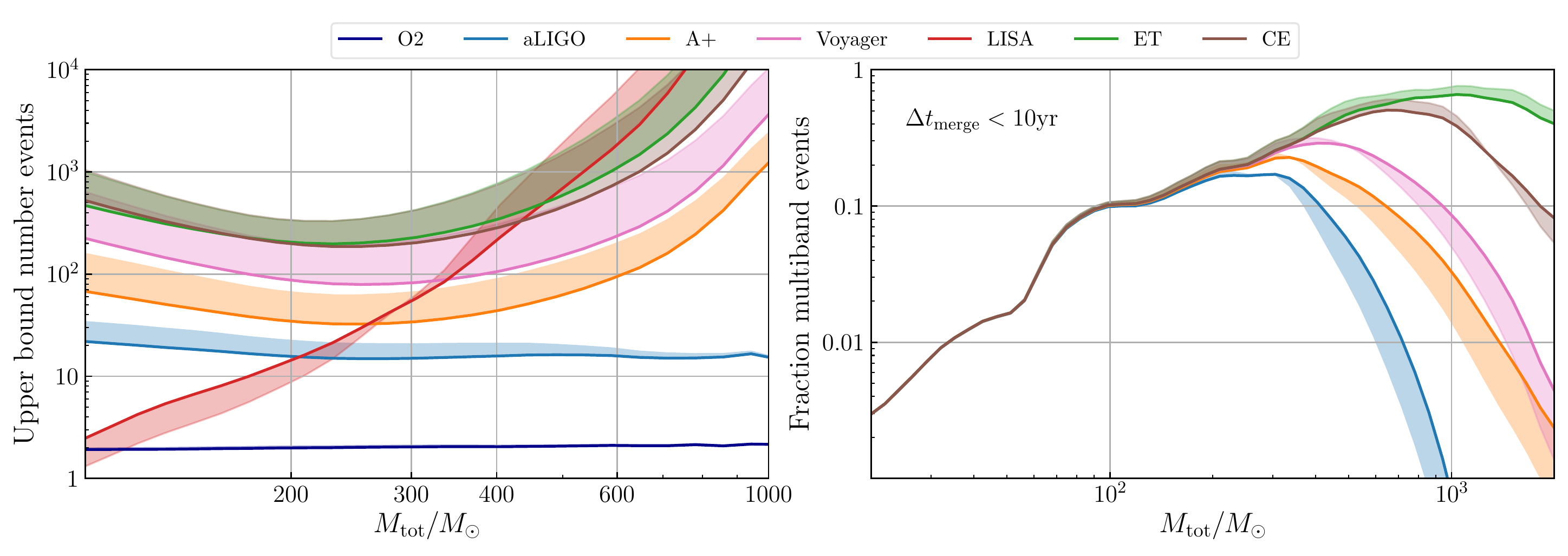}
\caption{(Left panel) Projection of the maximum number of events detected per year for ground-based detectors (except O2 which is fixed to 9 months) and in 4 years for {\it LISA}\/ given the upper bounds from the O1 and O2 runs. (Right panel) Fraction of multi-band events, defined as those {\it LISA}\/ detections merging within 10 years and being detected by a ground-based detector. Solid lines represent a constant merger rate with redshift, while the shaded areas delineate the difference resulting from a redshift evolution tracking the star formation rate.
}
\label{fig:nevents}
\end{figure*}

If we take as an input present upper bounds from O1-O2 (solid blue line in Fig.~\ref{fig:upper_bound_rate}),
we can extend current LVC analyses \citep{Salemi:2019ovz} and estimate the maximum number of detections above the PISN gap in future runs, cf. left panel of Fig.~\ref{fig:nevents}.
In one year of O4 at design sensitivity (aLIGO curve) there will be no more than 20 events across the mass range of interest.
Still, due to the weak constraints at $\Mtot>300\Msun$, there could be hundreds of events above these masses in one year of observation of O5 at A+ sensitivity. The width of the shaded regions in Fig.~\ref{fig:nevents} encapsulates the differences between a constant merger rate and one that evolves with redshift following the SFR~\citep{2018ApJ...863L..41F}. For example, we find that the A+ band is more pronounced than the aLIGO curve; this is due to its farther horizon, and therefore the larger number of relative detections.

\section{{Sensitivity beyond LIGO-Virgo}}
Far-side binaries will also be probed by future detectors. We will focus on the space-based mission {\it LISA}~\citep{2017arXiv170200786A}, and a third generation (3G) of ground-based detectors~\citep{3Gdetectors}.

{\it LISA}\/ provides an interesting perspective in the quest for binaries above the PISN mass gap. Instead of detecting the final stages of the merger and ring-down, as ground-based detectors do, it will be mostly sensitive to the inspiral. In fact, {\it LISA}\/ could detect BBHs which are still hundreds of years from merging, and monitor them during its entire observing lifetime. For this reason it is convenient to express the number of events that {\it LISA}\/ will detect as a function of detector frame frequency $f_d$:
\be
\frac{\dd^4N_\text{det}}{\dd m_1\dd m_2\dd z\dd f_d}=(1+z)\R(z)\frac{\dd V_c}{\dd z}\frac{\dd t_s}{\dd f_s}p(m_1,m_2)p_\text{det}\,,
\ee
where the time to coalesce in the source frame is computed assuming a circular orbit.
Differently to ground-based detectors, $p_\detc$ will be sensitive to the time variation of {\it LISA}'s antenna pattern over the course of observation of a given source.

Although the number of detections of different observatories is subject to the intrinsic merger rate, the ratio between them is independent of its local value $\R_0$, offering a way to leverage the redshift evolution of $\R(z)$.
Taking O2 upper bound as an input for $\R_0$, we see in Fig.~\ref{fig:nevents} that {\it LISA}\/ could detect in 4 years tens to hundreds of events for $\Mtot$ between $200$--$400\Msun$.
A difference between the ground-based prediction and {\it LISA}\/ observations could signal for instance that binaries have non negligible eccentricities.

 {\it LISA}\/ will also provide the opportunity of detecting the same GW event across different frequencies.
Stellar-mass BBHs below the PISN gap have been proposed as multi-band sources \citep{Sesana:2016ljz}, although {\it LISA}\/ high frequency sensitivity limits their number \citep{Moore:2019pke}.
If present in nature, IMBHs would be more promising candidates~\citep{Amaro_Seoane_2010,Jani:2019ffg,Sedda:2019aa}.
We present in the right panel of Fig.~\ref{fig:nevents} the \emph{fraction of multi-band events}, defined as the subset of {\it LISA}\/ detections that will merge within 10 years and be detected by a ground-based instrument.
Interestingly, the multi-band fraction peaks where the upper end of the PISN mass gap is expected to be found. In agreement with \citet{Gerosa_2019}, we find that for $\Mtot\lesssim100\Msun$, there is no difference for the multi-band ratio between 2G and 3G detectors.
On the contrary, for $\Mtot>200\Msun$ the difference among ground-based detectors are sizable.

We consider three concepts to follow aLIGO as 3G detectors: Voyager \citep{voyager}, Einstein Telescope (ET) \citep{Punturo_2010}  and Cosmic Explorer (CE) \citep{Abbott_2017}.
From the left panel of Fig.~\ref{fig:nevents} we find that 3G interferometers will detect more BBHs in 1 year than {\it LISA}\/  in 4 years for $\Mtot\lesssim400\Msun$, with up to several hundred events in this mass range. Moreover, since 3G detectors probe to higher redshifts, they are more sensitive to the merger rate redshift evolution. Finally, Fig.~\ref{fig:nevents} shows that ET/CE will be highly complementary to {\it LISA}\/ in terms of the fraction of multi-band events for the most massive BBHs.

\section{{Science with far-side binaries}}
There would be tremendous scientific potential associated with the detection of a new population of BBHs located on the far side of the PISN gap, including the nuclear physics of the PISN, the astrophysical modeling of formation channels, tests of gravity, and cosmological probes.
Here we focus on three concrete questions: \emph{(a)} how well could we constrain the upper end of the PISN gap (or, as we refer to it here, the minimum far-side mass), \emph{(b)} how can we use these BBHs to constrain the cosmic expansion history, and \emph{(c)} what would be the impact of the far side population on the background of unresolved binaries?

\begin{figure*}[t!]
\centering
\includegraphics[width = 0.98\textwidth]{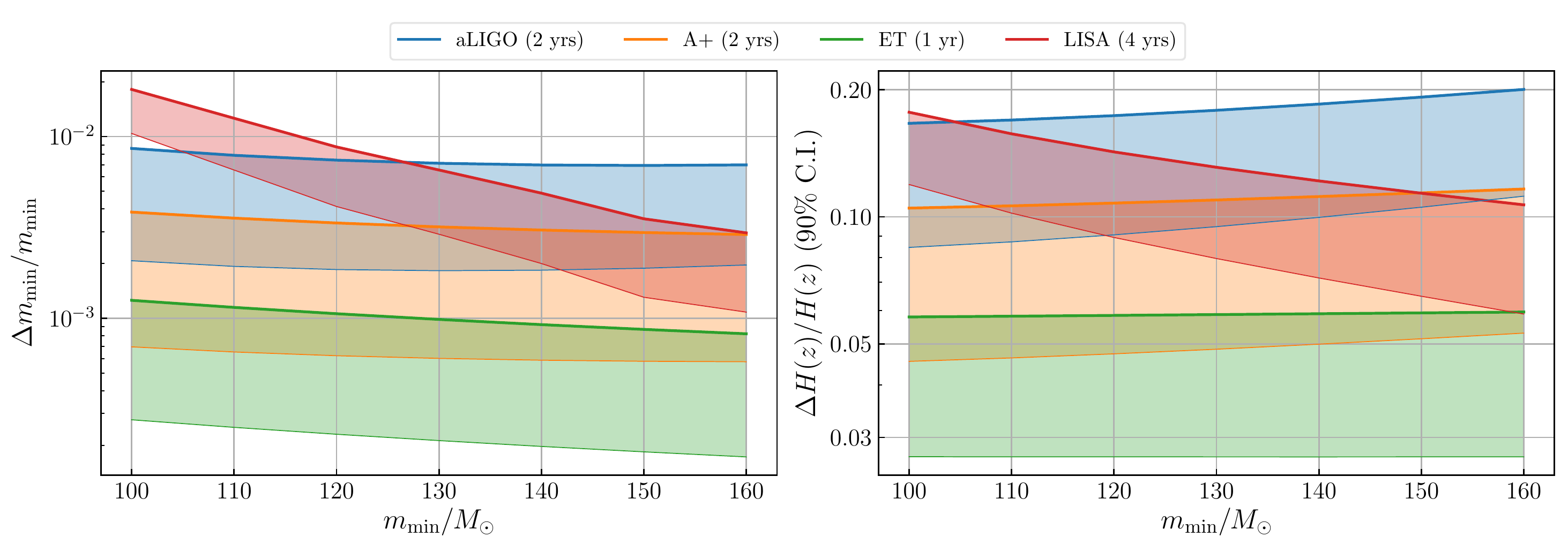}
 \caption{Estimated fractional error on the minimum mass of the population of BBHs above the PISN mass gap (left) and on the Hubble parameter $H(z)$ at $90\%$ confidence interval (C.I.) obtained from standardizable GW sirens (right). We assumed a uniform distribution masses from $\mmin$ to $\mmin+60 M_\odot$ with comoving merger rate $\R_c = 0.1\,\Gpc^{-3}\yr^{-1}$. The shaded regions represent the uncertainty in the redshift evolution of the merger rate between a constant rate (thick line) and a rate following the star formation rate (thin line).}
\label{fig:error_mmin}
\end{figure*}

\subsection{Minimum far-side mass}
\label{sec:pop}

In order to determine how well we can measure the minimum far-side mass, we need to know how well we determine the edges of a distribution given a finite number of random draws. We utilize the maximum separation estimation technique, which maximizes the geometric mean of the separations of data in the cumulative distribution function \citep{10.2307/2345411}.
 \footnote{We thank Maya Fishbach for bringing this to our attention, and for providing a solution to this problem.}

Since we are interested in the minimum mass of the BBH above the gap, it is convenient to work with the distribution of secondary masses, $m_2$ (by definition $m_2\leq m_1$). Assuming that the distribution of observed masses follows a power-law $p(m_2^\obs)\sim(m_2^\obs)^\alpha$ with $\alpha<-1$, then after $N$ events the error in our estimate of the minimum mass would be
\be \label{eq:mmin_error}
\frac{\Delta\mmin}{\mmin}\sim\left\vert\frac{1}{N(\alpha+1)}\right\vert\,.
\ee
Noticeably, this error scales faster than the typical $1/\sqrt{N}$ scaling; it is easier to find the edge of a distribution than the peak of a distribution. More details on the derivation of (\ref{eq:mmin_error}) can be found in Appendix \ref{app:determining_edge}.

We model our fiducial far-side population as a uniform distribution in primary mass and mass ratio with $\R_0=0.1\,\Gpc^{-3}\yr^{-1}$, and a fixed mass range $\mmax-\mmin$. This way, when we vary $\mmin$ in a given interval, for example $[100\Msun,160\Msun]$, the overall rate is preserved. Moreover, we have chosen the local merger rate to be in agreement with O2 upper bounds.

Taking as input the number of events for our fiducial far-side model and the slope of the distribution of detected secondary masses~\citep{2020ApJ...891L..27F}, we can estimate the error in the minimum mass from Eq.~\ref{eq:mmin_error}.
As shown in Fig.~\ref{fig:error_mmin}, the minimum mass could be precisely measured, to better than 1\% in most cases. This measurement is improved if
$\R(z)$ follows the SFR (c.f., the bottom of the colored bands), since this increases the number of detected binaries.
Since we are determining the minimum mass from $p(m_2^\obs)$, our results are subject to the assumption that mass ratios are uniformly distributed. If we were to determine $\mmin$ from the primary mass, which could be thought as a proxy for equal mass binaries, the errors $\Delta\mmin/\mmin$ would be a factor of 5--10 larger.

Two important assumptions are relevant in this estimate. First, we are assuming a unique population of far-side binaries. In principle, there could be multiple formation channels contributing binaries both within and above the gap, for instance through generations of successive mergers of lower-mass black holes.
It is not generally expected that such hierarchical mergers would result in a significant population at these large masses, although the precise rates of first-second and second-second generation mergers are model dependent. 
In what follows we make the generous assumption that far side BBHs exist at the highest rates consistent with current observations, while interloper binaries just below the upper edge of the PISN gap are suppressed (as compared to the population above the gap) as expected from theory.
As we quantify in Appendix \ref{app:in_gap_pop}, if the population of far-side binaries and in-gap binaries are comparable on both sides of the upper edge of the PISN gap, the inference on $\Delta m_\text{min}$ will be significantly degraded.
Second, we assume that the power-law of $p(m_2^\obs)$ is known. In practice, one would have to simultaneously determine $\alpha$ and $\mmin$, similar to the analysis in~\citep{Fishbach:2019ckx}. In this sense our estimates may be considered optimistic, although they should constitute a good approximation for sufficient numbers of detections.

\subsection{GW standard sirens}

Binary GW sources provide a direct measurement of the luminosity distance, and thereby can be used as standard sirens~\citep{Schutz:1986gp,Holz:2005df}.
In general, however, GW observations do not provide any direct information about the redshift of the sources, which is required to constrain the cosmic expansion history.
Redshifts can be obtained observing an electromagnetic counterpart~\citep{Holz:2005df,2006PhRvD..74f3006D,Nissanke:2013fka}, which was spectacularly accomplished with GW170817~\citep{Abbott:2017xzu}  and could lead to precision cosmological constraints in the near future~\citep{Chen:2017rfc,DiValentino:2018jbh}.
In the absence of counterparts, one can perform a statistical analysis ~\citep{Schutz:1986gp,DelPozzo:2011yh}, as has been applied to LIGO/Virgo detections~\citep{Fishbach:2018gjp,Soares_Santos_2019}, and in the future may be applied to {\it LISA}\/~\citep{Del_Pozzo_2018,PhysRevD.95.083525}.

Alternatively, one could use features in the mass distribution to directly calibrate the population. This is because in the detector frame we observe redshifted masses $m^z_{1,2}=(1+z)m_{1,2}$. Therefore, if we know the source-frame mass distribution, we can infer the redshift distribution of the population and use the GW events as \emph{standardizable sirens}.
In this respect, the PISN gap is a particularly promising feature. \citet{Farr:2019twy} proposes using the lower edge to constrain $H(z)$ to the percent level at $z\sim0.8$. In what follows we pursue a similar approach, applying it to the opposite (upper) end of the mass gap.

Working with the distribution of detector frame masses, we can obtain the minimum mass as in the previous section.
Since we have shown that $\mmin$ can be well constrained, the individual error in $H(z)$ is going to be dominated by the measurement uncertainty in $d_L$, which is subject to the detector's calibration and the degeneracy with inclination. We are interested in the number of detected events with information about the minimum mass per luminosity distance bin, $N_{\Delta \mmin}$, which we quantify by $m^\detc_2-2\sigma_{m_2}<\mmin^{d_L}$. The error in the determination of $H(z)$ can then be approximated by
\be \label{eq:errorHz}
\frac{\Delta H(z)}{H(z)}\sim\frac{\Delta d_L(z)/d_L(z)}{\sqrt{N_{\Delta\mmin}}}\,.
\ee
This estimation will be most precise at the peak of the redshift distribution of detected GWs.

We present our estimates for the measurement of the cosmological expansion rate in the right panel of Fig.~\ref{fig:error_mmin}. A+, {\it LISA}, and ET could constrain $H(z)$ to better than $10\%$ at $90\%$ confidence interval, with ET potentially achieving $<5\%$ if the merger rate follows the SFR (similar results hold for CE).
With two years of observations of aLIGO, the limit will remain larger than $10\%$. Of course, these numbers could improve if the observing time or the merger rate increases.
Interestingly, because the peak of the detected redshift distribution is different for each detector and minimum far-side mass, this method is sensitive to the expansion rate at different cosmic times. In particular, {\it LISA}\/ could provide a better measurement of the local expansion rate $H_0$,  aLIGO of the equation of state of the dark energy, and ET/CE of the dark matter.
Although these constraints are less sensitive than other standard siren tests, they provide an entirely independent determination, and could be used to improve the global GW constraint.
Moreover, far-side GW sirens could allow {\it LISA} to cross-calibrate with LIGO-Virgo's low redshift standard sirens~\citep{Schutz:1986gp,Holz:2005df,2006PhRvD..74f3006D}.

We emphasize that Eq.~\ref{eq:errorHz} is a rough, order of magnitude estimate.
Once data is collected one should perform a proper Bayesian analysis as in \citet{Farr:2019twy}.
Moreover, this test assumes that the PISN gap does not change across cosmic history, which could introduce an important systematic uncertainty.
Recent analyses have shown however that  both the lower end of the gap \citep{Farmer_2019} and its width \citep{2020arXiv200606678F} are robust against ambient factors and nuclear reaction rates.
Interestingly, one could then use the location of the lower end, which soon will be precisely measured \citep{Farr:2019twy}, to calibrate the upper end of the gap. Additionally, systematic errors from lensing could be relevant for 3G detectors~\citep{1998PhRvD..58f3501H,2005ApJ...631..678H,2010PhRvD..81l4046H}.

Finally, a population of interloper BBHs below the upper edge of the PISN gap would also affect the inference on $H(z)$, increasing the error in $\Delta \mmin$ and decreasing the number of events with information about $\mmin$. Unless the edge of the gap is completely erased from the mass distribution, e.g. if the fraction of interloper binaries $f$ is equal to that of far sides ($f=0.5$), the errors on $\Delta H(z)$ will remain dominated by the luminosity distance measurement error. We can then estimate the effect of the  interloper population within the gap by computing the change in $N_{\Delta\mmin}$ in Eq. \ref{eq:errorHz} (see Appendix~\ref{app:in_gap_pop} for details).
For example, if the number of events just below the upper edge of the gap (i.e., interloper BBHs) is 1/3 of the number of events above the edge (i.e. far side BBHs), $f=1/3$, and we find that $\Delta H(z)/H(z)$ will be $\approx 1/\sqrt{1-2f} = 1.7$ times larger.

\begin{figure}[t!]
\centering
\includegraphics[width = 0.95\columnwidth]{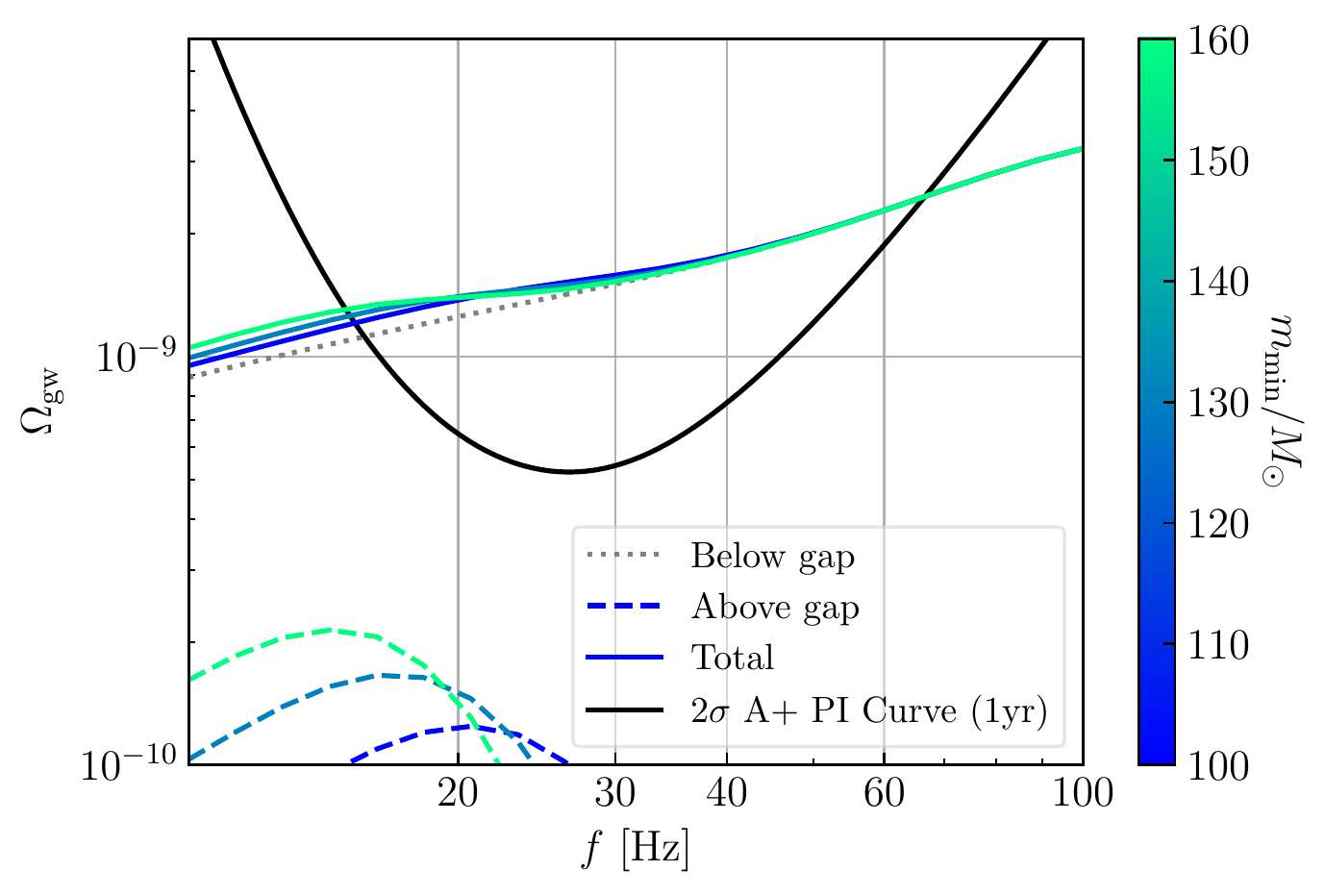}
 \caption{Impact of the population above the PISN mass gap on the energy density spectrum $\Ogw$ of the unresolved BBHs during 1 year of A+, as a function of the minimum far-side mass $\mmin$. For reference we include the $2\sigma$ power-law integrated (PI) curve.}
 \label{fig:sgwb}
\end{figure}

\subsection{Stochastic GW Background}

Binaries above the PISN mass gap will also contribute to the stochastic GW background (SGWB) of unresolved binaries.
We study the SGWB produced by our fiducial model above the gap $\Ogw^\text{above}$, together with a population below the gap consistent with observations \citep{LIGOScientific:2018jsj} leading to $\Ogw^\text{below}$.
We focus on ground-based detector's band since at LISA frequencies both will add a similar background \citep{Mangiagli_2019}.

As shown in Fig.~\ref{fig:sgwb}, $\Ogw^\text{above}$ peaks at the frequencies where LVC detectors are most sensitive (given by the $2\sigma$ power-law integrated curve \citep{PhysRevD.88.124032}).
In 1 year of A+, the bump in the total SGWB induced by the additional far-side BHs lies well within the sensitivity. In fact, for this particular population, the total SNR \citep{PhysRevD.59.102001} will be  $\sim6$, with the relative difference with and without $\Ogw^\text{above}$ being $10\%$ below 25~Hz. This hints that this characteristic breaking of the standard $\Ogw\sim f^{2/3}$ scaling could be detected at $1\sigma$ within $\sim4$ years (recall that SNR $\propto\sqrt{T_\text{obs}}$).
Note that neutron-star and neutron-star-black-hole binaries do not affect the spectral shape of $\Ogw$ below 100 Hz \citep{10.1093/mnras/stt207,PhysRevLett.120.091101,2019arXiv190811073C}.

The detectability of this distinctive spectral feature is subject to the ratio $\Ogw^\text{above}/\Ogw^\text{below}$, which depends on the merger rate of each population. Moreover, this spectral distortion could be degenerate with certain models of redshift evolution \citep[e.g., see Fig. 2 of][]{PhysRevLett.116.131102}. 
However, given the connection of the breaking of the power-law with $\mmin$, holistic analysis of resolved and unresolved GWs \citep[e.g.][]{PhysRevX.8.021019,2020arXiv200312152C,2020arXiv200409700S,2020arXiv200412999S} could be used to disentangle the end of the PISN gap.

\section{{Conclusions}} \label{sec:conclusions}
We have shown that a population of far side, post PISN gap binary black holes is a promising target
for both ground- and space-based detectors. Detection of these far side BBHs would provide a wealth of astrophysical and cosmological information. 
Far-side binaries would constitute the most massive sources detectable by LIGO/Virgo, and are a primary target for current IMBH searches \citep{Virgo:2012aa,Aasi:2014iwa,Salemi:2019ovz}. We find that present upper bounds \citep{Salemi:2019ovz} allow for up to several tens of far-side detections during O4 and O5. In the absence of detections their merger rate would be strongly constrained, to less than $0.01\,\yr^{-1}\Gpc^{-3}$ by the end of O5.
We demonstrate that {\it LISA}\/ could also see this population, complementing stellar-mass LIGO/Virgo+LISA binaries \citep{Sesana:2016ljz,Moore:2019pke}, and, in fact, far-side BBHs might dominate the fraction of multi-band events.
Additionally, we show that the minimum far-side mass could be used to ``standardize'' GW standard sirens, enabling direct constraints on $H(z)$ at redshift $\sim0.4$, $0.8$, and $1.5$ with {\it LISA}, aLIGO, and ET respectively. We consider the impact of a possible population of interloper black holes within the PISN mass gap, and provide a formalism for assessing their impact on PISN standard siren cosmology.
Finally, we show that the upper edge of the PISN gap may also leave a distinctive imprint on the stochastic background of unresolved sources.
Far-side binaries could also be an important target population for a deci-hertz observatory~\citep{Sedda:2019aa}, and would open new tests of gravity with standard sirens \citep{Ezquiaga:2018btd,Lagos:2019aa}, and multi-band events \citep{2020arXiv200612137D}.
Future observations will either uncover a population of far-side black holes, or provide strong limits on its existence.

\acknowledgments
We are grateful to Maya Fishbach for her critical insights on determining the minimum mass of the population above the PISN mass gap.
We are also thankful for the important constructive feedback from two anonymous referees, as well as to Sylvia Biscoveanu, Christopher Berry, Reed Essick, Amanda Farah, and Xingjiang Zhu for their comments on the manuscript.
JME is supported by NASA through the NASA Hubble Fellowship grant HST-HF2-51435.001-A awarded by the Space Telescope Science Institute, which is operated by the Association of Universities for Research in Astronomy, Inc., for NASA, under contract NAS5-26555. DEH is supported by NSF grants PHY-1708081 and PHY-2011997, and also gratefully acknowledges support from a Marion and Stuart Rice Award.
Both authors are also supported by the Kavli Institute for Cosmological Physics through an endowment from the Kavli Foundation and its founder Fred Kavli.

\appendix

\section{{Methods}}
\label{app:methods}
In the following we provide further details on our methodology. We describe the observing scenarios considered, population specifications, and our methodologies for incorporating  GW detection efficiencies, measurement errors, and selection biases. We use pyCBC \citep{alex_nitz_2019_3546372} with the IMRPhenomD approximant \citep{PhysRevD.93.044006} to compute the waveform of non-spinning BBHs.
We set the threshold signal to noise ratio (SNR) for detection in a single detector at 8, and assume a {\it Planck}\/ 2018 cosmology \citep{Collaboration:2018aa}.

\paragraph{Observing scenarios:} \label{app:observing_scenarios}

We consider Advanced LIGO and Virgo runs following the latest version of~\citet{Aasi:2013wya} (specifically LIGO public document P1200087-v58 of early 2020). For O1/O2/O3 we consider 116/269/365 days of observation with 41/46/60$\%$ coincident operation of both aLIGO detectors. For O4 and O5 we adopt 2 years of observation at design sensitivity and 2 years at the upgraded design (A+) with a 70$\%$ coincident operation time. We use the sensitivity curves described in \citet{Aasi:2013wya}, which can be found at \citet{sensitivity_curves_ligo}.

For third generation detectors, Voyager, Einstein Telescope and Cosmic Explorer, we adopt the sensitivity curves given in \citet{sensitivity_curves_3g}. Finally, for the future space-based detector {\it LISA}\/ we use the sensitivity curve defined in \citet{Cornish:2018dyw}, which can be downloaded from GitHub \citep{lisa_tools}.

\begin{figure*}[t!]
\centering
\includegraphics[width = 0.98\textwidth]{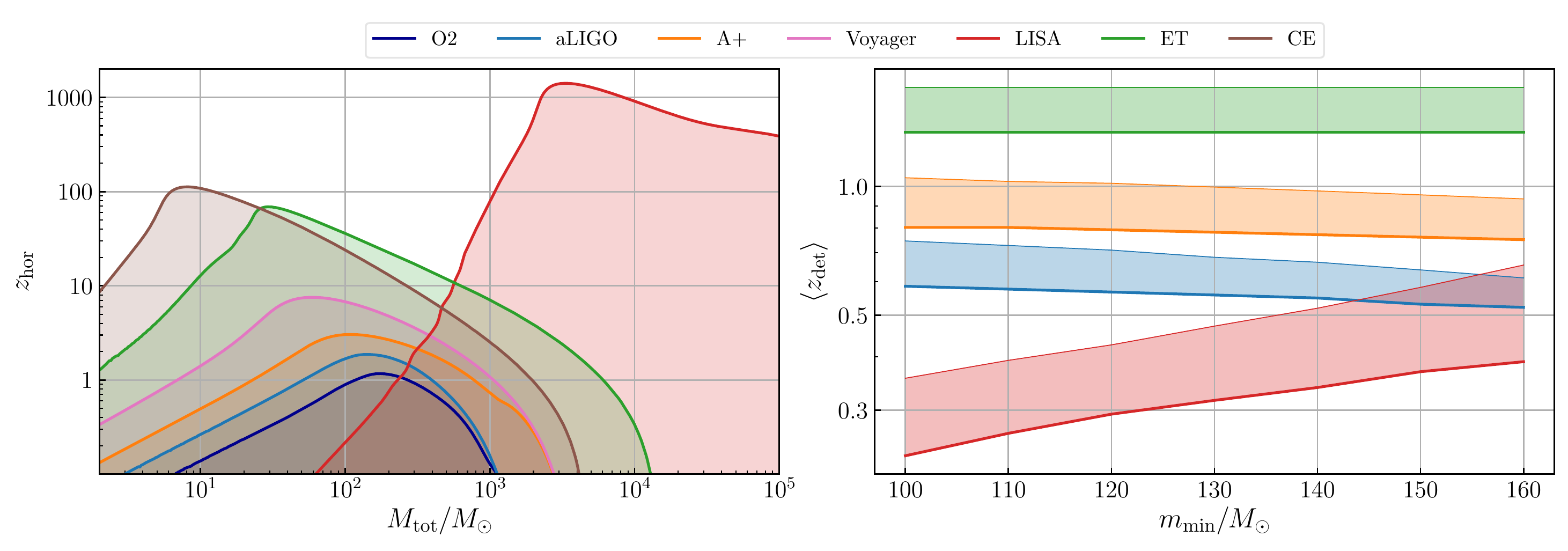}
 \caption{On the left, horizon redshift as a function of the total source frame mass for a signal-to-noise ratio detection threshold $\rho_\text{th}=8$. We assume 4 years of {\it LISA}\/ observations. On the right, most probable detected redshift for our fiducial population of BBHs above the PISN mass gap.}
 \label{fig:zhor}
\end{figure*}

\paragraph{Sky localization sensitivity:}

In order to determine the probability of detecting a GW from a given binary system, defined as $p_\detc$ in the main text, we take into account  the sky position, orientation, and inclination angle. For ground-based detectors, since their antenna pattern is basically fixed during the detection time, we use the cumulative distribution function $p_\detc(w)$ of having a signal-to-noise ratio (SNR) above a given threshold $\rho_{\text{th}}$ given an optimal SNR $\rho_\text{opt}$ (face-on, over-head), where $w=\rho_\text{th}/\rho_\text{opt}$ \citep{Chen:2017wpg}. A table with $p_\detc(w)$ values can be found in \citet{distance_tool}. Given the uncertainty of the actual configuration of future networks, in our analysis we focus on single detector SNR and set $\rho_\text{th}=8$. We use this criteria when estimating the upper bound on the comoving merger rate and find consistency with  the LVC-IMBH analyses \citep{Salemi:2019ovz}, although there un-modeled searches of short duration signals are also performed.

For {\it LISA}\/ the situation is more complicated because the SNR of a single detection is accumulated over a longer period of time, and must take into account a time-dependent antenna pattern. We use the tools developed in \citet{Cornish:2018dyw}, which provide the {\it LISA}\/ sensitivity for any sky location and inclination, and facilitate the computation of {\it LISA}'s antenna pattern cumulative distribution function for different masses and initial frequencies, $p_\detc(w,\Mchirp,f_i)$. We note that for binaries that are observed early in their inspiral, and thereby stay in band over the whole {\it LISA}\/ mission, the effect of the sky localization is small since it tends to average out. However, the SNR of binaries that are going to merge or leave the {\it LISA}\/ band  within the {\it LISA}\/ observational window can be significantly affected by the position in the sky and inclination.

The sensitivity of a detector can also be described by its horizon distance. In Fig. \ref{fig:zhor} we summarize the different detector configurations used throughout the analysis.

\paragraph{Binary black-hole populations:}

In the main text we study two representative scenarios for the redshift evolution of the merger rate $\R(z)$: \emph{(i)} a constant merger rate, and \emph{(ii)} a merger rate following the star formation rate (SFR). For the evolving merger rate, we adopt a redshift dependence following Eq.~15 of \citet{Madau:2014bja}, normalized to $\R_0$ at $z=0$, and without the inclusion of a time delay distribution.

We model our fiducial far-side population as a uniform distribution in primary mass and mass ratio with $\R_0=0.1\,\Gpc^{-3}\yr^{-1}$, and a fixed mass range $\mmax-\mmin$. With this choice when we vary $\mmin$ in a given interval, for example $[100\Msun,160\Msun]$, the overall rate is preserved. In addition, the local merger rate is chosen to be in agreement with O2 upper bounds.

For the population below the mass gap studied in the stochastic background analysis, we specifically consider a power-law distribution between $5$ and $42\Msun$ with slope $1.6$, $\R_0=30\Gpc^{-3}\yr^{-3}$, and following the SFR.

\paragraph{Mock GWs:}
In order to estimate the measurement errors and selection biases, we follow the prescription of \citet{Fishbach:2019ckx} (explained in detail in their appendix A).
We begin by assigning a measurement error to the observed SNR
\be
\rho_\obs=\mathcal{N}(\rho,1)\,,
\ee
which we assume to follow a normal distributions $\mathcal{N}$ centered at the true value.
Then, we compute the typical error in the observed redshifted chirp mass $\Mz^\obs$, symmetric mass ratio $\eta^\obs$ and angular projection term $w^\obs$. Again, we assume that the observed quantities follow normal distributions $\mathcal{N}$ centered at their true values and with a variance scaling inversely with the observed SNR $\rho_\obs$, namely we draw the observed values from the following distributions
 \begin{align}
\log\Mz^\obs&=\mathcal{N}\lb\log\Mz,\sigma_{\log\Mz}\cdot \rho_\text{th}/\rho\rb\,, \label{eq:error_logMc}\\
 \eta_\obs &= \mathcal{N}\lb\eta,\sigma_\eta\cdot \rho_\text{th}/\rho\rb\,, \label{eq:error_eta}\\
 w_\obs &= \mathcal{N}\lb w,\sigma_w\cdot \rho_\text{th}/\rho\rb\,, \label{eq:error_w}
 \end{align}
 where $0\leq\eta_\obs\leq1/4$ and $0\leq w_\obs\leq1$ must be imposed. Finally, the uncertainty in the observed masses $\sigma_{ m^\obs_{1,2}}$ and luminosity distance $d_L^\obs$ can be directly drawn from the above assumptions. One should note that this simulates the maximum likelihood rather than the full posterior of the parameters \citep{Fishbach:2019ckx}.

Our choices for the measurement uncertainty for each detector are summarized in Table 1. We make these choices in order to recover the typical errors in the masses and luminosity distance to be expected from detailed parameter estimation analyses of 2G detectors \citep{Vitale_2017}, 3G detectors \citep{PhysRevD.95.064052}, and {\it LISA}\/ \citep{2020arXiv200300357M}.
 We emphasize that these errors are subject to the precise networks of detectors active during the time of observation.

Knowing the typical errors of the observed masses, $m^\obs_{1,2}$, allows us to estimate the uncertainty in our determination of the minimum mass of the distribution (see Fig. 3). We use this information in order to estimate the error in the determination of the expansion rate in Eq. (5) of the main text.

\begin{table*}[!]
\centering
\setlength{\tabcolsep}{6pt}
\renewcommand{\arraystretch}{0.9}
\begin{tabular}{c | c | c | c || c | c}
\hline
\hline
Detector & $\sigma_{\log\Mz}$ & $\sigma_\eta$ & $\sigma_w$ & $\Delta m^\obs_{1,2}/m^\obs_{1,2}$ ($90\%$ C.I.) & $\Delta d^\obs_L/d^\obs_L$ ($90\%$ C.I.) \\ [0.3ex]
\hline\hline
O4 (aLIGO) & $8\cdot10^{-2}$ & $1\cdot10^{-2}$ & $8\cdot10^{-2}$ & $40\%$ & $50\%$\\
O5 (A+) & $3\cdot10^{-2}$ & $5\cdot10^{-3}$ & $5\cdot10^{-2}$ & $25\%$ & $40\%$ \\
Voyager& $1\cdot10^{-2}$ & $2\cdot10^{-3}$ & $5\cdot10^{-2}$ & $20\%$ & $40\%$ \\
{\it LISA}\/ & $1\cdot10^{-5}$ & $1\cdot10^{-3}$ & $3\cdot10^{-2}$ & $5\%$  & $25\%$ \\
ET & $5\cdot10^{-3}$ & $7\cdot10^{-4}$ & $2\cdot10^{-2}$ & $10\%$ & $20\%$ \\
CE & $5\cdot10^{-2}$ & $7\cdot10^{-4}$ & $2\cdot10^{-2}$ & $10\%$ & $20\%$ \\
\hline
\hline
\end{tabular}
 \label{tab:errors}
\caption{Summary of the measurement errors used for each detector and the consequent typical $90\%$ confidence interval (C.I.) for the observed, source frame masses $m_{1,2}^\obs$ and luminosity distances $d_L^\obs$ of threshold events. We assume that the observed values of the logarithm of the redshifted chirp mass $\log\Mz^\obs$, symmetric mass ratio $\eta^\obs$ and angular projection term $w^\obs$ follow Gaussian distributions with variances at threshold SNR $\rho_\text{th}=8$ given by each column, cf. Eqs. (\ref{eq:error_logMc}-\ref{eq:error_w}).}
\vspace{-0.2cm}
\end{table*}

\section{{Determining the edges of a distribution}}
\label{app:determining_edge}
Under the assumption that the average separation of the observed events is larger than the measurement uncertainty for each event, we can use the maximum separation estimation technique \citep{10.2307/2345411} to asses the question of how well the edge of a distribution can be constrained. This technique allows us to estimate the error in the intrinsic parameters of a distribution given $N$ random draws by maximizing the geometric mean of spacings in the cumulative distribution function of the data.

For example, for $N$ observed systems taken from a uniform distribution, with a minimum, $m_1$, and a maximum, $m_N$, observed mass, then the error in the estimate of the minimum mass of the distribution $\Delta{m_\text{min}}=m_1-m_\text{min}^\text{obs}$ and the maximum mass $\Delta{m_\text{max}}=m_\text{max}^\text{obs}-m_N$ are given by
\be
\Delta{m_\text{min}}=\frac{m_N-m_1}{N-1}=\Delta{m_\text{max}}\,.
\ee
As expected, the larger the number of events which are detected close to the edge in the mass distribution, the better the constraint on the location of the edge. This could be considered as a continuum extension of the `serial number analysis' in statistics \citep{10.2307/2281038}.
In the case that the observed distribution follows a power-law, $p(m^\text{obs})\propto (m^\text{obs})^\alpha/((m_\text{max}^\text{obs})^{\alpha+1}-(m_\text{max}^\text{obs})^{\alpha+1})$, the error in minimum and maximum mass are given by
\be \label{eq:mmin_error_general}
\Delta{m_\text{min}}=\frac{m_1(N-1)^{\frac{1}{\alpha+1}}-(m_1^{\alpha+1}N-m_N^{\alpha+1})^{\frac{1}{\alpha+1}}}{(N-1)^{\frac{1}{\alpha+1}}}
\ee
and
\be \label{eq:mmax_error_general}
\Delta{m_\text{max}}=\frac{(m_N^{\alpha+1}N-m_1^{\alpha+1})^{\frac{1}{\alpha+1}}-m_N(N-1)^{\frac{1}{\alpha+1}}}{(N-1)^{\frac{1}{\alpha+1}}}\,.
\ee
As before, we find that additional events close to the edge tend to do a better job of constraining the location of the edge, but now the value of the slope of the distribution can impact the measurement. For example, a value of $\alpha<-1$ improves the measurement of the minimum mass, while $\alpha>1$ improves the measurement of the maximum mass.
Focusing on the minimum mass determination, if $m_N^{\alpha+1}\ll m_1^{\alpha+1}N$ (limit valid when $\alpha\ll-1$ and/or $N\gg1$), we can expand Eq. (\ref{eq:mmin_error_general}) in $N\gg1$ to arrive at the result in the main text of Eq.~(4). Note also that we have to make sure that the range of masses, namely $m_N-m_1$, is sufficiently large for the average separation of the events to be larger than the typical error in the observed mass, which we can calculate using the procedure described in the previous section.

\section{{Impact of an interloper population below the upper edge of the PISN gap}}
\label{app:in_gap_pop}

\begin{figure}[t!]
\centering
\includegraphics[width = 0.45\columnwidth]{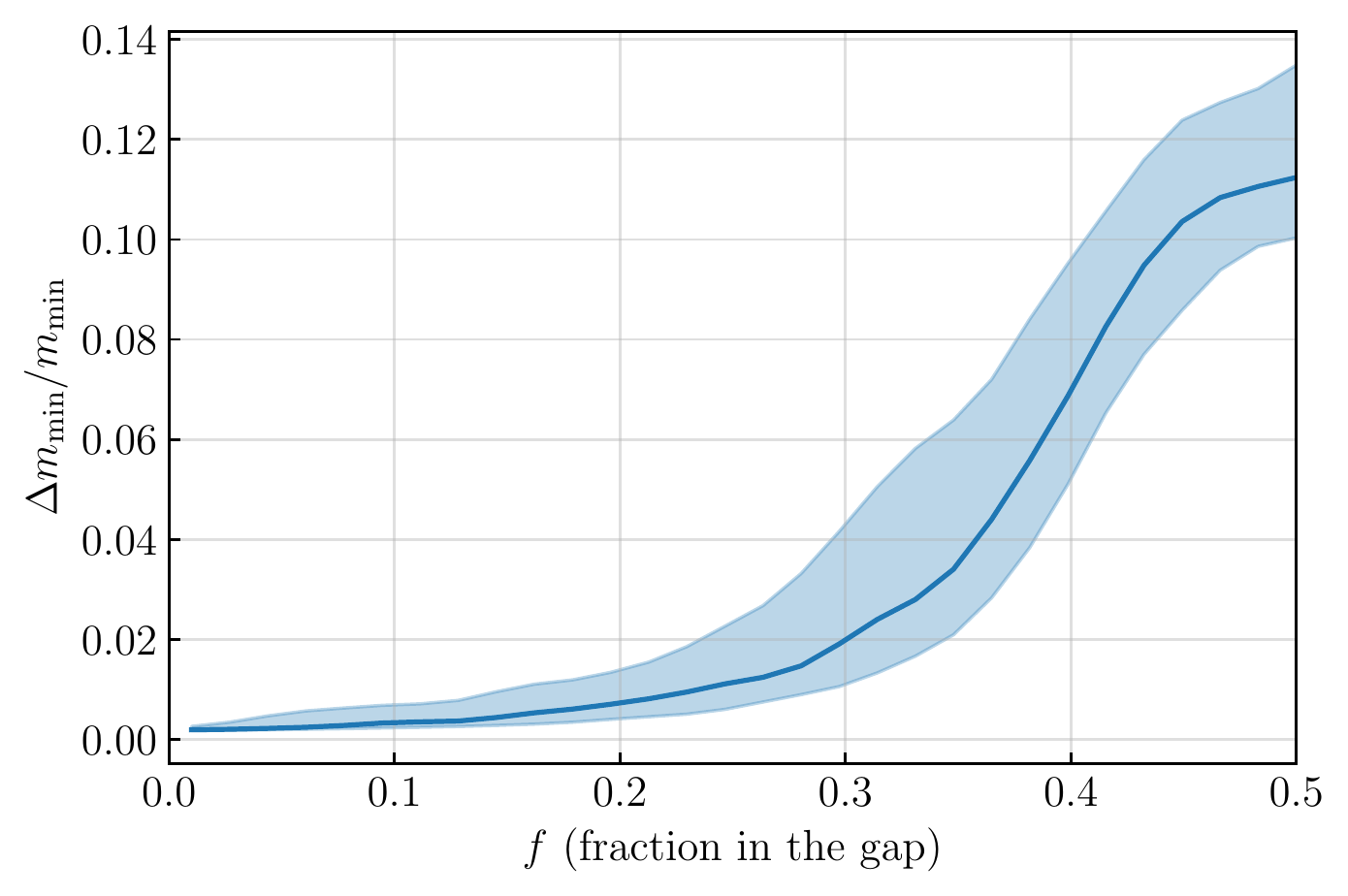}
 \caption{Impact on the inference of the minimum far-side mass $\mmin$ (or upper edge of the PISN gap) of the fraction of binaries in the PISN gap $f$. We consider a 3G-like scenario with 100 detections. The solid line corresponds to the mean relative error and the bands indicate the $1\sigma$ deviation after 100 simulations.
We choose uninformative priors for $\mmin$ and $f$.}
 \label{fig:pop_in_gap}
\end{figure}

In the main text we have worked under the assumption that the population of BBHs in the PISN gap, interloper binaries, is negligible compared to the population of far-side binaries. It is conceivable, however, that these populations contribute similarly to the observed rate of events or even that the population above the gap is more suppressed. In order to quantify this ``pollution" of the upper edge of the PISN gap, we concentrate on the events just above and below the edge.
We consider a simple toy model in which both the populations on either side of the edge are uniform distributions, and their relative rate is controlled by the fraction $f$ (where $0<f<1$):
\begin{equation}
p(m_1)=\begin{cases}
    f/\Delta m_\text{gap}, & \text{if $\mmin-\Delta m_\text{gap}<m_1<\mmin$}.\\
    (1-f)/\Delta m_\text{above}, & \text{if $\mmin<m_1<\mmin+\Delta m_\text{above}$}\\
    0, & \text{otherwise}.
  \end{cases}
\end{equation}
We further fix $\Delta m_\text{gap}=\Delta m_\text{above}=20\Msun$ to make $f$ the only parameter quantifying the density of events in the vicinity of $\mmin$. 
Using this model as input, we determine how the inference of $m_\text{min}$ is affected by $f$.
For concreteness we consider a 3G-like scenario with hundreds of detections. 
This adds the additional simplification that the intrinsic and detected density of events per mass bin will be approximately the same.
The results are displayed in Fig. \ref{fig:pop_in_gap} where the intuitive guess of the degradation of the inference of $\mmin$ is quantified for this particular example.
The error in $\mmin$ is maximum when the density of events above and below the edge is the same, i.e. when $f=0.5$.
At this point of equality the inference in $\mmin$ is lost completely since there is no longer any feature to find! The error is then just determined by the priors, which we set to be uniform in $\mmin$ between $50$ and $160\Msun$ and in $f$ between 0 and 1.
However, as shown in the plot, a precise inference of $\mmin$ is still possible away from $f\sim0.5$.\footnote{In principle $\mmin$ can also be determined when $f>0.5$ since there is still a feature in the mass distribution, although this situation is not expected from theory.}
We thus conclude that if the density of interloper events just {\em below}\/ the upper edge of the gap is comparable to the density of far-side binaries ($f\sim 0.5$), the relative error of the upper edge of the PISN gap could be significantly larger than when the population in the gap is subdominant ($f\ll1$) as was assumed in the main text.
A similar contamination would occur for the other cases considered in Fig. \ref{fig:error_mmin}.

A simple estimate of the additional error induced by the interloper population below the upper edge of PISN gap is possible. One can assume that the $f\cdot N$ events from the population in the gap act as a background. From this perspective, if we remove this background and only consider the ``extra'' binaries above the edge, we have a similar situation to that in Eq. \ref{eq:mmin_error} and Appendix \ref{app:determining_edge}, finding an inverse scaling with the number of events with information about the edge which is now $(1-2f)N$.  The error is thus simply scaled by $\sim1/(1-2f)$.

Similarly, a population in the gap would also affect the role of far-side binaries as standard sirens. When the fraction in the gap is far from $0.5$, the most relevant effect for the determination of $H(z)$ would be the decrease in the number of events with information about the upper edge of the PISN gap,  $N_{\Delta\mmin}$ in Eq. \ref{eq:errorHz}. This is because the measurement uncertainty in $d_L$ is still larger than $\Delta\mmin$, cf. Figs. \ref{fig:error_mmin}, \ref{fig:pop_in_gap} and Table 1. 
Following the previous argument, when the far-side population dominates, we can approximate $N^{\text{tot}}_{\Delta\mmin}\approx (1-2f)N^\text{above}_{\Delta\mmin}$, making the relative error in $H(z)$ increase by $\sim1/\sqrt{1-2f}$. 
In the opposite regime when $f\sim 0.5$, the error in $H(z)$ is dominated by the error in redshift which comes directly from $\Delta \mmin$. Accordingly, for these particular scenarios the errors in the determination of the edge compromise cosmological tests.

\bibliography{gw_refs}
\bibliographystyle{aasjournal}
\end{document}